\documentclass[prb,superscriptaddress,twocolumn]{revtex4-2}
\usepackage{graphicx}
\usepackage{times,amsmath}
\usepackage{epsfig}
\usepackage{color}
\usepackage{graphicx}% Include figure files
\usepackage{dcolumn}% Align table columns on decimal point
\usepackage{bm}% bold math
\usepackage{bookmark}
\usepackage{tabularx}% bold math
\usepackage{hyperref}
\usepackage{multirow}
\usepackage{array,mathtools,amssymb,booktabs,makecell}
\hypersetup{colorlinks=true, citecolor=blue, filecolor=blue, linkcolor=blue, urlcolor=blue}

\usepackage{float}
\usepackage{adjustbox}
\usepackage{upgreek}
\usepackage{soul}
\usepackage[utf8]{inputenc}
\usepackage[T1]{fontenc}
\usepackage[shorthands=off, % Don't translate quotation marks followed by vocals to umlauts
            ngerman         % Neue deutsche Rechtschreibung
]{babel}
\usepackage[autostyle]{csquotes}
\MakeOuterQuote{"}

\renewcommand{\selectlanguage}[1]{}

\makeatletter
\let\saved@includegraphics\includegraphics
\AtBeginDocument{\let\includegraphics\saved@includegraphics}
\renewenvironment*{figure}{\@float{figure}}{\end@float}
\makeatother

\begin{document}

%\title{Signatures of Hundness in UN}
%\title{Signatures of Hundness in the spin-orbital coupled $f$-electron system UN}
\title{Hund's physics extends to actinide $f$ electron systems}

\author{Byungkyun Kang}
%\email[]{bkkang@utep.edu}
\affiliation{Department of Physics 500 W University Ave, The University of Texas at El Paso, El Paso, Texas 79968, USA}

\author{Roy N. Herrera-Navarro}
%\email[]{bkkang@utep.edu}
\affiliation{Department of Physics 500 W University Ave, The University of Texas at El Paso, El Paso, Texas 79968, USA}

\author{Stephen S. Micklo}
%\email[]{bkkang@utep.edu}
\affiliation{Department of Physics 500 W University Ave, The University of Texas at El Paso, El Paso, Texas 79968, USA}

\author{Mark R. Pederson}
\affiliation{Department of Physics 500 W University Ave, The University of Texas at El Paso, El Paso, Texas 79968, USA}

\author{Eunja Kim}
\email[]{ekim4@utep.edu}
\affiliation{Department of Physics 500 W University Ave, The University of Texas at El Paso, El Paso, Texas 79968, USA}

%\author{Mark R. Pederson}
%\affiliation{Department of Physics 500 W University Ave, The University of Texas at El Paso, El Paso, Texas 79968, USA}

\begin{abstract}

Uranium 5$f$ electrons often yield heavy-fermion behavior via Kondo screening. However, the pronounced bad-metallic transport of uranium mononitride (UN) defies an incoherent Kondo explanation. Using density-functional theory combined with dynamical mean-field theory, we show that UN is a strongly correlated bad metal. The dominant correlations arise from intra-atomic Hund’s exchange interaction between two 5$f$ electrons, which aligns local magnetic moments and produces large quasiparticle mass renormalization. This identifies UN as a 5$f$-electron analogue of a Hund’s metal—a paradigm chiefly associated with transition-metal $d$ systems. Our results motivate a re-examination of the interplay between Mott, Kondo, and Hund-driven correlations across actinide correlated materials.

\end{abstract}

\maketitle

\textit{Introduction.} Strong electron correlations underpin a broad spectrum of emergent phenomena in condensed matter, including unconventional superconductivity, local-moment magnetism, heavy-fermion/Kondo physics, and incoherent bad-metal transport \cite{anderson_science1987,scalapino_rmp2012,alan_jmmm2003,khuntia_jmmm2019,kondo_ptp1964,stewart_rmp1984,gunnarsson_rmp2003}. One pathway to bad-metallic normal-state behavior is proximity to a Mott insulator—Mottness—where strong on-site Coulomb repulsion drives incoherence, as exemplified by the high-Tc cuprates \cite{phillips_rpp2009,basov_rmp2005,masatoshi_rmp1998}, the canonical Mott system V$_2$O$_3$ \cite{limelette_sci2003}, VO$_2$ \cite{qazilbash_sci2007}, and rare-earth nickelates RNiO$_3$ (R = rare-earth element) \cite{maria_jpcm1997}.
In multiorbital systems, a distinct correlation mechanism arises from on-site Hund’s exchange interaction $J_\textrm{H}$, which suppresses interorbital charge fluctuations and promotes incoherence even away from a Mott transition. This Hundness drives bad-metal behavior in the normal state of iron-based superconductors \cite{kamihara_iron-based_2006,kamihara_iron-based_2008,de_medici_janus-faced_2011,de_medici_hunds_2011,georges_strong_2013,isidori_charge_2019}. The resulting Hund’s metals \cite{haule_coherenceincoherence_2009,yin_kinetic_2011,chibani2021lattice} provide a paradigmatic reference for understanding correlation effects in iron-based superconductors \cite{georges_strong_2013,haule_coherenceincoherence_2009,yin_kinetic_2011,de_medici_hunds_2017,lanata_orbital_2013,villar_arribi_hund-enhanced_2018,bascones_orbital_2012,ryee_nonlocal_2020} and in ruthenates \cite{georges_strong_2013,werner_spin_2008,mravlje_coherence-incoherence_2011,hoshino_superconductivity_2015,mravlje_thermopower_2016}.
While Hundness has been primarily established in transition-metal $d$-electron materials, density functional theory combined with dynamical mean-field theory (DFT+DMFT) identifies uranium mononitride (UN) as a bad metal in which the electrical resistivity reflects not only electron–phonon scattering but also a substantial electron–electron contribution \cite{quan_prb2011}. This observation motivates a systematic examination of correlation mechanisms in $f$ electron systems, and particularly assessing whether its bad-metal behavior is governed by Mottness, Kondo hybridization, or Hundness.

Uranium-based compounds host a wealth of correlated-electron phenomena arising from the dual itinerant–localized character of 5$f$ electrons, strong spin–orbit coupling, and hybridization with conduction bands. Notably, these characteristics make several uranium compounds prime candidates for spin-triplet superconductivity~\cite{Aoki2019review}. Prototypical examples include the ferromagnetic superconductors URhGe ($T_\textrm{c}=$ 0.25 K) ~\cite{Aoki2001} and UCoGe ($T_\textrm{c}=$ 0.8 K) ~\cite{Huy2007}, where superconductivity emerges from a ferromagnetically ordered state, supporting the scenario of equal-spin pairing ~\cite{Aoki2019review}. More recently, superconductivity with $T_\textrm{c} \approx 2$ K was discovered in Curie-paramagnetic UTe$_2$ ~\cite{Ran2019science,Rosa2021,Aoki2021review}, which is one of the highest critical temperatures among uranium-based superconductors.

These unconventional superconducting and magnetic ground states are widely believed to originate from strong electronic correlations, including Kondo lattice physics, in proximity to magnetic quantum criticality. Quantum critical fluctuations commonly produce non-Fermi-liquid behavior over extended temperature ranges in heavy-fermion materials. Canonical uranium systems that exhibit the coexistence of Kondo behavior and ferromagnetic order include UTe~\cite{schoenes_jlcm1986}, UCu$_{0.9}$Sb$_2$~\cite{bukowski_jac2005}, and UCo$_{0.5}$Sb$_2$~\cite{bukowski_jac2005,bukowski_pss2006} , with relatively high Curie temperatures $T_\textrm{C}=$ 102 K, 113 K, and 64.5 K, respectively. Kondo physics is likewise prominent in Pauli-paramagnetic UFe$_2$Si$_2$, where X-ray spectroscopy reveals the dual local atomic-like multiplet character of the U-5$f$ states ~\cite{andrea_pnas2020}, consistent with earlier reports of Kondo-like behavior at low temperatures ~\cite{szytula_jmmm1988}. In the Kondo-lattice ferromagnet USbTe, a large anomalous Hall conductivity has been attributed to intrinsic Berry curvature arising from Kondo hybridization between U-5$f$ moments and conduction electrons ~\cite{byung_usbte}.

In many uranium compounds, heavy-fermion behavior is well described by a Kondo lattice picture in which hybridization between local U-5$f$ moments and conduction electrons drives coherence. By contrast, the U-5$f$ electronic character of UN remains contentious. Angle resolved photoemission spectroscopy reports predominantly itinerant U-5$f$ states in UN~\cite{shin_prb2012}, whereas nuclear magnetic resonance reveals features of a dense Kondo system above $T_\textrm{N}$ ~\cite{ogloblichev_prb2021}. Complementary DFT+DMFT calculations have suggested non-Fermi-liquid responses and bad-metallic transport arising from strong electron correlations~\cite{quan_prb2011}.

Here we investigate the electronic structure of UN in its paramagnetic phase. We find sizable U-5$f$ magnetic moments and strong electron correlations. Analysis of the density of states, the self-energies, and U-5$f$ multiplet configurations points to the correlations dominated by Hund’s coupling (“Hund physics”), rather than a conventional Kondo-lattice mechanism. This identifies UN as a Hund-correlated 5$f$ metal.

\begin{figure}[ht]
\centering
\includegraphics[width=0.5
\textwidth]{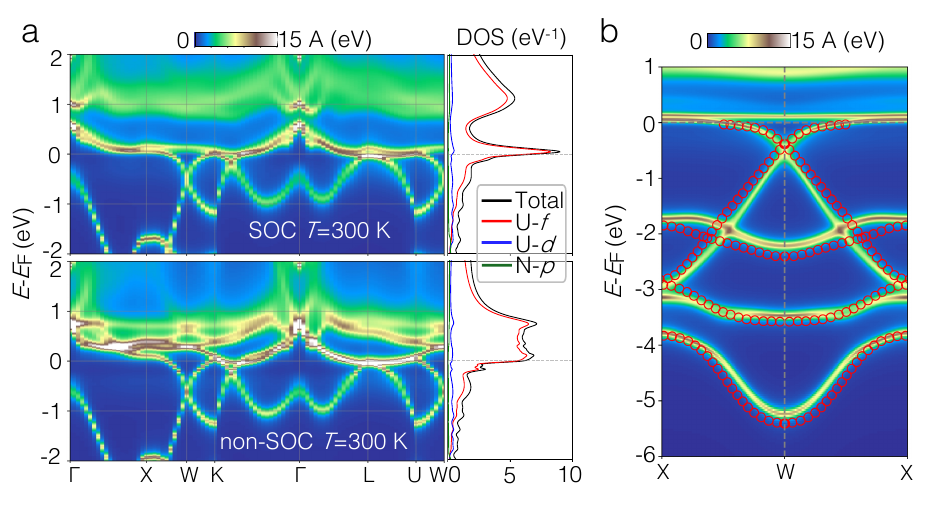}
\caption{\label{Fig_spectral}\
\textbf{Electronic structure of UN.}
\textbf{a}, Calculated spectral function and density of states (DOS) without and with inclusion of SOC at 300 K.
\textbf{b}, Calculated spectral function at $T$ = 75 K, along the X-W-X high symmetry line. ARPES measurements at $T$ = 75 K~\cite{shin_prb2012} are denoted by red circles.
}
\end{figure}

\textit{Electronic structure of UN.} We adopted the experimental lattice constant $a = 4.89~\textrm{\AA}$ for UN, which crystallizes in the space group $F\bar{4}3m$ (No. 216)~\cite{knott_prb1980}. We employed the constrained random-phase approximation (cRPA)~\cite{aryasetiawan2004frequency}, as implemented within the linearized quasiparticle self-consistent GW plus dynamical mean-field theory (LQSGW+DMFT) framework~\cite{choi2019comdmft}, to determine the on-site Coulomb interaction $U_{\textrm{C}}$ and the Hund’s exchange interaction $J_{\textrm{H}}$ for the U-5$f$ manifold in UN. The resulting $U_{\textrm{C}}=1.87$ eV agrees well with the previously predicted $U_{\textrm{C}}=1.85$ eV, which correctly describes the ground state and electronic structure of UN within density functional theory~\cite{jian_jap2013}; it is also consistent with $U_{\textrm{C}}=2.0$ eV, which reproduces the experimental ordered moment of $1.7~\mu_{\textrm{B}}$ in LDA+$U$ calculations~\cite{sun_mrs2014}, and with $U_{\textrm{C}}=2.2$ eV, which yields the experimental cell volume~\cite{lukoyanov_eps2015}. Nevertheless, our $U_{\textrm{C}}$ is smaller than the value $U_{\textrm{C}}=5.2$ eV obtained from fully self-consistent many-body GW calculations~\cite{quan_prb2011,kutepov_prb2010}. The computed Hund’s exchange interaction, $J_{\textrm{H}}=0.59$ eV, is in excellent agreement with the value reported by Ogasawara et al.~\cite{ogasawara_prb1991}, $J_{\textrm{H}}=0.58$ eV for U-5$f$ derived from Slater integrals for the trivalent U$^{3+}$ ion.

Using the values of $U_{\textrm{C}}$, $J_{\textrm{H}}$, and the experimental lattice parameter, we employed charge-self-consistent DFT+DMFT as implemented in the COMSUITE package~\cite{choi2019comdmft} to obtain the electronic structure of UN. All calculations were performed with spin-orbit coupling (SOC) unless otherwise specified. Figure~\ref{Fig_spectral}a shows the spectral functions and density of states (DOS), with and without SOC, at $T=300$ K. The calculated U-5$f$ occupancy of 2.40 (with SOC) and 2.35 (without SOC) is consistent with the measured value of $2.2 \pm 0.5 e$~\cite{norton_prb1980}. As shown in Fig.\ref{Fig_spectral}b, the calculated spectral function along the high-symmetry line X--W--X at 75 K is in good agreement with angle-resolved photoemission spectroscopy (ARPES) measured at the same temperature in its paramagnetic phase~\cite{shin_prb2012}. The pronounced peak just below the Fermi level, attributed to substantial U-5$f$ weight in the angle-integrated photoemission spectra~\cite{shin_prb2012}, is likewise captured by the calculated DOS.

\begin{figure}[ht]
\centering
\includegraphics[width=0.5
\textwidth]{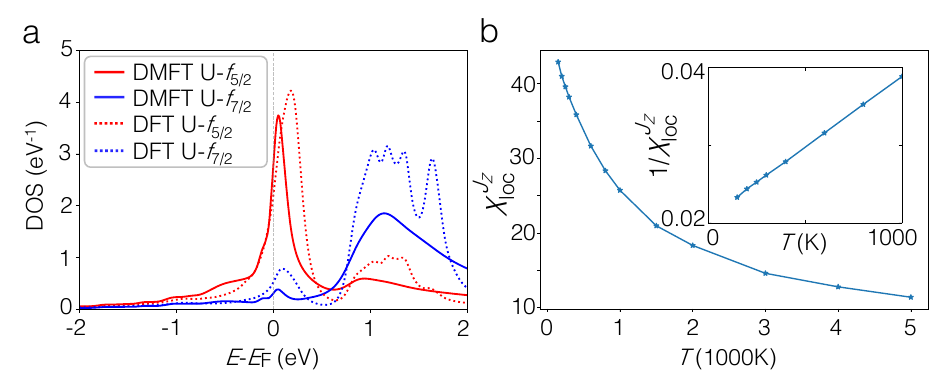}
\caption{\label{Fig_fdos_sus}\
\textbf{Electron correlations in UN.}
\textbf{a} Density of states of U-5$f$ multiplets projected within DFT+DMFT (solid lines) and DFT (dotted lines).
\textbf{b} Calculated local total angular momentum susceptibility. Inset show inverse of the calculated local total angular momentum susceptibility.
}
\end{figure}

As shown in Fig.\ref{Fig_spectral}a, without SOC the broad metallic bands in the vicinity of the Fermi level are dominated by U-5$f$ character. With SOC, the U-5$f$ manifold is split into $j=5/2$ and $j=7/2$ multiplets. The unoccupied part of the spectral function associated with the $j=7/2$ multiplet forms a broad, incoherent feature centered around 1 eV above the Fermi level, whereas the $j=5/2$ multiplet exhibits a sharp peak pinned at the Fermi level. This sharp DOS centered at the Fermi level is consistent with a Hund-driven metallic state. In a Hund’s metal, the upper and lower Hubbard bands substantially overlap, and the spectrum is dominated by an incoherent feature with enhanced weight at the Fermi level~\cite{kang_nqm2023,deng2019signatures,stadler_prb2021,stadler_annphy2019,stadler_prl2015}.

\begin{table*}[]
\caption{Electron occupations (occ.), relative crystal-field energies ($\Delta_E$) of the U-5$f$ levels obtained by projecting the mean-field Hamiltonian onto the U-5$f$ subspace~\cite{choi2019comdmft}, and quasiparticle renormalization factors $Z$ for the U-5$f$ orbitals. At 600 K, the crystal-field energies and orbital occupations are essentially unchanged relative to those at 100 K.}\label{table_occ}
%\tiny
\scriptsize
\begin{center}
\scalebox{1}{
\begin{ruledtabular}
\vspace*{5mm}
%\scalebox{1.0}{
\begin{tabular}{c|ccccccc|cccccccc}

   & \multicolumn{1}{c|}{$j$} & \multicolumn{6}{c|}{5/2} & \multicolumn{8}{c}{7/2}  \\
   \hline
   & \multicolumn{1}{c|}{$j_{z}$} &-2.5& -1.5 & -0.5 & 0.5 & 1.5 & 2.5 & -3.5 & -2.5 & -1.5 & -0.5 & 0.5 & 1.5 & 2.5 & 3.5 \\

  \hline
  \multirow{3}{*}{$T$=100 K} & \multicolumn{1}{c|}{occ.} & 0.28 & 0.31 & 0.28 & 0.28 & 0.31 & 0.28 & 0.08 &0.09 & 0.08 & 0.08 & 0.08 & 0.08 & 0.09 & 0.08 \\
  & \multicolumn{1}{c|}{$\Delta_E$}  & 0.12 & 0.0 & 0.15 & 0.15 & 0.0 & 0.12 & 1.02 &0.84 & 0.93 & 1.03 & 1.03 & 0.93 & 0.84 & 1.02 \\
  & \multicolumn{1}{c|}{$Z$}  & 0.57 & 0.59 & 0.59 & 0.59 & 0.59 & 0.57 & 0.67 & 0.82& 0.82 & 0.59 & 0.59 & 0.82 & 0.82 & 0.69 \\  
  \hline

  \multirow{1}{*}{$T$=600 K} 
  & \multicolumn{1}{c|}{$Z$}  & 0.80 & 0.78 & 0.79 & 0.79 & 0.78 & 0.80 & 0.87 & 0.84& 0.82 & 0.89 & 0.89 & 0.82 & 0.84 & 0.87 \\

  %\hline
\end{tabular}
\end{ruledtabular}}
\end{center}
\end{table*}

\textit{Strong electron correlations in UN.}
In a Hund’s metal, electron correlations are primarily driven by Hund physics~\cite{stadler_prb2021}, in contrast to a Mott metal where Mott physics dominates. Both mechanisms generate substantial electron correlations. To assess the presence of electron correlations in UN, we compare the DOS obtained from DFT+DMFT and DFT, as shown in Fig.~\ref{Fig_fdos_sus}a. The comparison reveals clear band renormalization within DFT+DMFT: the bandwidth of the $j=5/2$ multiplet at the Fermi level is reduced relative to DFT, implying enhanced effective mass, while the $j=7/2$ multiplet is more broadened than in DFT, indicative of significant incoherent scattering.

The quasiparticle renormalization factor, defined from the derivative of the imaginary part of the self-energy on the Matsubara axis,
$Z=\left[1-\left.\frac{\partial\mathrm{Im}\Sigma(i\omega)}{\partial(i\omega)}\right|_{i\omega\to 0^{+}}\right]^{-1}$\cite{marianetti_prl2008,bo_ncomm2021,arsenault_prb2012}, is shown in Table~\ref{table_occ}.
The associated mass enhancement due to electron correlations is given by $m^{\star}/m=1/Z$~\cite{marianetti_prl2008,bo_ncomm2021}.
As an indicator of correlation strength, $Z$ ranges from unity (weak correlations) to zero (strong correlations).
At high temperature, the U-5$f$ orbitals exhibit large $Z$ values, indicative of weak correlations.
Upon cooling to 100 K, $Z$ for the partially occupied U-5$f$ states in the $j=5/2$ multiplet decreases, signaling enhanced electronic correlations and an increased quasiparticle effective mass of the U-5$f$ electrons.

\begin{figure}[ht]
\centering
\includegraphics[width=0.3
\textwidth]{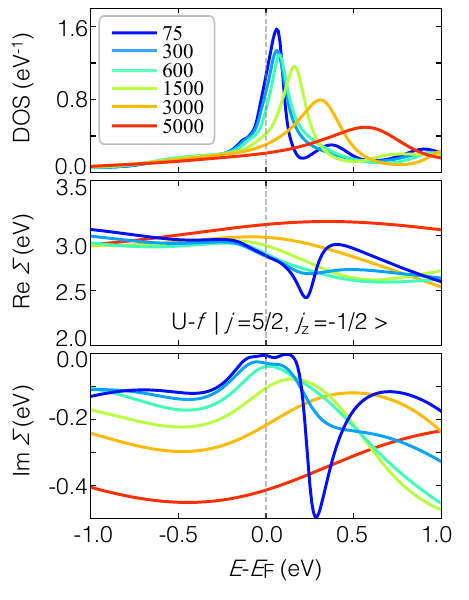}
\caption{\label{Fig_self}\
\textbf{Hundness electronic structure.}
Projected density of states, the real part and imaginary part of the self-energy of the U-5$f$ state of the $|j= 5/2,j_{z}= 1/2\rangle$. The temperature unit is Kelvin.
}
\end{figure}

The local total angular-momentum susceptibility was evaluated as
$\chi_{\mathrm{loc}}^{J_z}=\int_0^\beta d\tau\langle J_z(\tau)J_z(0)\rangle$.
As shown in Fig.\ref{Fig_fdos_sus}b, $\chi_{\mathrm{loc}}^{J_z}$ of U-5$f$ states in UN exhibits Curie–Weiss behavior. This is consistent with experiments reporting Curie–Weiss temperature dependence of the paramagnetic susceptibility of single-crystal UN in the [100], [110], and [111] magnetic field directions~\cite{doorn_jltp1977,ogloblichev_prb2021}, indicative of substantial local magnetic moments typically arising from strong electron correlations. Together, these results demonstrate that strong electron correlations persist in UN and are consistent with previous DFT+DMFT work identifying a large mass enhancement in UN~\cite{quan_prb2011}.

\textit{Signature of Hundness in UN.} To elucidate the origin of the strong electron correlations, we computed the U-5$f$ self-energy in UN. Figure~\ref{Fig_self} presents the orbital-resolved DOS and self-energy for the $|j=5/2, j_{z}=-1/2\rangle$ state. The DOS exhibits a pronounced temperature dependence. Notably, up to $T=5000$ K there is no indication of pseudogap formation— a hallmark of Mottness wherein the high-temperature spectral weight at the Fermi level is depleted due to quasiparticle spectral-weight transfer, resulting in a pseudogap at the Fermi level~\cite{deng2019signatures}. Instead, the $|j=5/2, j_{z}=-1/2\rangle$ DOS follows the behavior expected for Hundness: at high temperature a single incoherent peak with substantial weight already exists at the Fermi level, and upon cooling a quasiparticle resonance develops out of this incoherent background~\cite{deng2019signatures}. Consistently, at $T=75$ and $300$ K the self-energy shows Hund’s metallic fingerprints, a shoulder-like structure in $\operatorname{Im}\Sigma(\omega)$ and an inverted slope of $\operatorname{Re}\Sigma(\omega)$ near $\omega=0$~\cite{stadler_annphy2019,stadler_prl2015}. Together, these observations point to Hund-driven, rather than Mott-driven, correlations in UN.

\begin{figure}[ht]
\centering
\includegraphics[width=0.5
\textwidth]{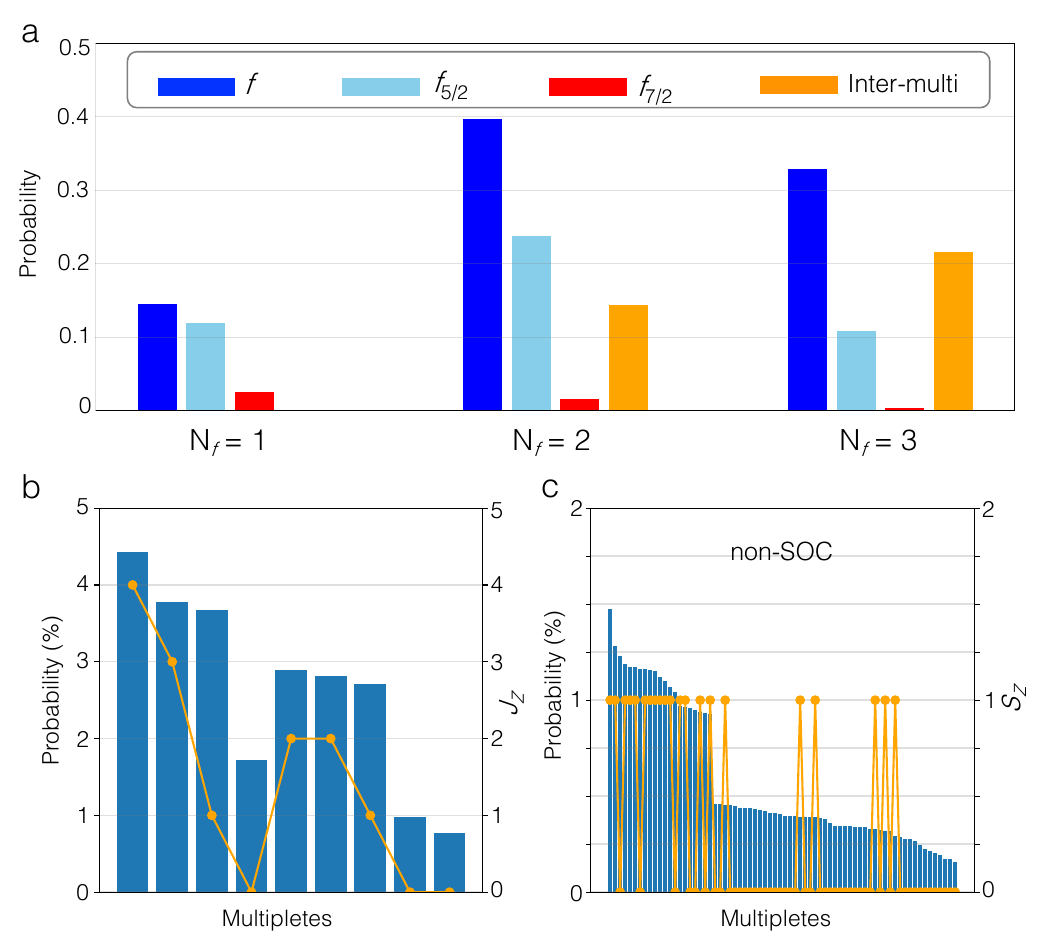}
\caption{\label{Fig_prob}\
\textbf{Valence histograms for the U-5$f$ multiplets.}
\textbf{a}, Probability distribution for the U-5$f$ multiplets in UN over atomic configurations, resolved by 5$f$ total charge $N_f$; it reports the fraction of time the impurity resides in each $N_f$ valence.
\textbf{b} Probabilities within the $N_f=2$ valence resolved by $j=5/2$ multiplets; each state is labeled by the total local angular-momentum projection $J_z$.
\textbf{c} Non-SOC counterpart of panel b: probabilities within the $N_f=2$ valence resolved by spin multiplets; each state is labeled by the total spin projection $S_z$.
}
\end{figure}

We investigated the presence of the Kondo effect in UN by analyzing the spectral functions at temperatures as low as 75 K. Our results did not reveal the characteristic kink-like band structure associated with Kondo hybridization between U-5$f$ and conduction electrons at the Fermi level. Such a band structure is a hallmark of the Kondo effect and is observed in uranium-based Kondo lattices such as UTe$_2$~\cite{byung_ute2,kang2023dual,kang_ute2fs} and USbTe~\cite{byung_usbte}, as well as in the rare-earth-based superconducting nickelate NdNiO$_2$~\cite{kang_ndnio2}. The absence of the Kondo effect in UN is further supported by the decline in electrical resistivity upon cooling~\cite{doorn_roh_jltp1997}, which lacks the characteristic logarithmic temperature dependence indicative of Kondo scattering  ($\rho$ $\sim$ -$\ln(T)$). Therefore, our analysis suggests that the observed electron correlations in UN down to 75 K are not attributable to the Kondo effect.

Stadler et al.\cite{stadler_prl2015,stadler_annphy2019} proposed that a defining feature of Hund metals is spin–orbital separation: the orbital degrees of freedom are quenched below an orbital Kondo scale $T_{K,\mathrm{orb}}$, which is typically much larger than the spin Kondo temperature $T_{K,\mathrm{spin}}$. This hierarchy of energy scales enables spin freezing at sufficiently low temperatures, once the orbital sector is inactive, leading to quasi-static local moments~\cite{philip_prl}. However, in the presence of strong spin–orbit coupling, spin is not a good quantum number. We therefore diagnose Hundness in UN via the total-angular-momentum channel within $j-j$ coupling scheme: the sizable $J_z$ fluctuations. 
%reflected by the Curie–Weiss-like $\chi_{\mathrm{loc}}^{J_z}$, together with the characteristic Hund’s-metal self-energy features, indicate that Hund’s coupling stabilizes large $J$-sector moments and governs the electronic correlations in UN.

Figure~\ref{Fig_prob}a shows the valence histogram for the U-5$f$ multiplets in UN. The histogram is constructed from the impurity’s reduced density matrix by performing a partial trace over all non‑5$f$ degrees of freedom and projecting onto the atomic 5$f$ eigenbasis. In this way it quantifies the generalized valence, wherein the 5$f$ shell in the solid fluctuates among a small set of atomic configurations and spends appreciable weight in each~\cite{shim_nature2007}.
The probability for U-5$f$ total charge $N_f=2$ is the largest among other charge configurations. However, in comparison to other uranium compounds, including UTe$_2$, USbTe, and USbSe~\cite{byung_ute2,kang2023dual}, UN exhibits more pronounced valence fluctuations with enhanced $N_f=3$ weight, indicating a more itinerant 5$f$ character, consistent with experimental observations of 5$f$ itinerancy in UN~\cite{shin_prb2012}. 
This further indicates that UN may lack sufficiently robust local magnetic moments to induce pronounced Kondo scattering of conduction electrons—unlike UTe$_2$\cite{kang_ute2fs,byung_ute2,Eo2021} and USbTe\cite{byung_usbte}—due to the itinerant 5$f$ character.
Nonetheless, the bad-metallic behavior in UN~\cite{quan_prb2011} can be attributed to strong electron correlations driven by Hund’s coupling. As shown in Fig.~\ref{Fig_prob}a, within the $N_f=2$ valence both 5$f$ electrons predominantly occupy the $j=5/2$ multiplet. Within the multiplet, the most probable two-electron states maximize $|J_z|$, corresponding to occupations with $m_j=\pm 5/2$ and $\pm 3/2$ that yield total $J_z=\pm 4$, as shown in Fig.~\ref{Fig_prob}b.
This is consistent with that Hund’s exchange interaction, $J_{\textrm{H}}$, gives rise to the lowest Coulomb interaction energy for the $f^2$ configuration at $|J_z|=4$ in the $j$–$j$ coupling scheme ~\cite{takashi_prb2003}.

Crystal-field splitting can enhance Hundness by reducing orbital symmetry in the three-orbital Hubbard model~\cite{fabian_prb2019}. In contrast, sufficiently strong crystal fields can promote Mottness, allowing the energy of spin singlet state to fall below the spin triplet state within a two-orbital model~\cite{philipp_prl2007}. As shown in Table~\ref{table_occ}, for UN the crystal field splitting within the $j=5/2$ multiplet is $\sim 0.15$ eV, which is sizable yet smaller than the computed Hund’s exchange interaction, $J_{\textrm{H}}=0.59$ eV. This moderate crystal-field splitting supports a Hund-dominated regime in UN. This conclusion is also consistent with the $f^2$ configuration in the $j$-$j$ coupling scheme with crystal field splitting, where the $J=4$ component has the largest weight in the ground-state wave function~\cite{takashi_prb2003}.

Although neglecting SOC is not realistic for UN, the non-SOC calculation yields spectral function and a U-5$f$ occupancy of 2.35 that are qualitatively similar to the SOC case (see Fig.\ref{Fig_spectral}a). To gain intuition, we examine the spin-state composition of the $N_f=2$ valence in the non-SOC model (see Fig.\ref{Fig_prob}c). The spin-triplet multiplets have substantially higher probability than the spin-singlet multiplets, indicating a strong intra-atomic Hund’s coupling even without SOC. This further supports the conclusion that Hundness governs the electron correlations in UN.

\textit{Conclusion.} We investigated the electronic structure of UN in its paramagnetic phase using charge–self–consistent DFT+DMFT with SOC and cRPA-derived interactions. Multiple, mutually consistent signatures indicate strong electron correlations governed by Hund physics: (i) Curie–Weiss behavior of the local total–angular–momentum susceptibility, evidencing sizable fluctuating local magnetic moments; (ii) the development upon cooling of a quasiparticle resonance at the Fermi level, emerging from a broad incoherent background, accompanied by mass renormalization; and (iii) valence histograms that favor high-$|J_z|$ configurations (or high-spin states in the non-SOC limit), characteristic of strong intra-atomic Hund’s coupling. Together, these results establish UN as a Hund’s metal.

Our findings highlight that Hund physics can play a critical role in actinide compounds—systems often interpreted primarily in terms of Mott or Kondo mechanisms—thus offering a fresh perspective on bad-metallic behavior in $f$-electron materials. They also motivate future studies of the interplay between Hund, Mott, and Kondo physics across the actinide and lanthanide families.

\section*{Acknowledgments}  
This research was performed using funding received from the U.S. Department of Energy, Office of Nuclear Energy’s Nuclear Energy University Program (NEUP).

\bigskip
\textbf{Competing Interests} The authors declare no competing interests.

\bigskip
\textbf{Data availability} The data that support the ﬁndings of this study are available from the corresponding
authors upon reasonable request.

\bigskip

\bibliography{ref}

\end{document}